\documentclass{emulateapj} 

\begin{document}

\newcommand{\ssfr}{{ $< -2.5 \log_{10}\ {\rm  Gyr^{-1}}$}}
\newcommand{\ha}{{ $< 2.5 EW(H\alpha)/\sigma(EW(H\alpha))$}}
\newcommand{\ngems}{171}
\newcommand{\ncosmos}{1161}
\newcommand{\nhiz}{1332}
\newcommand{\nsdss}{33086}
\newcommand{\blah}{$2\sigma$}
\newcommand{\dfrac}{54 $\pm$ 6\%}
\newcommand{\lfrac}{70 $\pm$ 8\%}

\newcommand{\etal}{{\em et~al.\,}}
\title{A constant limiting mass scale for flat early-type galaxies from z=1 to z=0: density evolves but
shapes do not.  \altaffilmark{1}}
\altaffiltext{1}{Based on observations with the NASA/ESA Hubble Space
Telescope, obtained at the Space Telescope Science Institute, which is
operated by the Association of Universities for Research in Astronomy,
Inc. under NASA contract No. NAS5-26555.}                                       

\author{Bradford P. Holden\altaffilmark{2}}
\author{Arjen van der Wel\altaffilmark{3}}
\author{Hans-Walter Rix\altaffilmark{3}}
\author{Marijn Franx\altaffilmark{4}}

\altaffiltext{2}{UCO/Lick Observatories, University of California,
  Santa Cruz, 95065, USA; holden@ucolick.org}
\altaffiltext{3}{Max-Planck Institute for Astronomy, K\"{o}nigstuhl 17, D-69117, Heidelberg, Germany; 
e-mail:vdwel@mpia.de,rix@mpia.de}
\altaffiltext{4}{Leiden Observatory, University of Leiden, P.O. Box 9513, 2300 R.A., Leiden, The Netherlands; 
e-mail: franx@strw.leidenuniv.nl}

\shorttitle{Shape Evolution in Early-types}

\begin{abstract} 
 We measure the evolution in the intrinsic shape distribution of
  early-type galaxies from $z\sim 1$ to $z\sim 0$ by analyzing their
  projected axis-ratio distributions. We extract a low-redshift sample
  ($0.04 < z < 0.08$) of early-type galaxies with very low
  star-formation rates from the SDSS, based on a color-color selection
  scheme and verified through the absence of emission lines in the
  spectra. The inferred intrinsic shape distribution of these
  early-type galaxies is strongly mass dependent: the typical
  short-to-long intrinsic axis-ratio of high-mass early-type galaxies
  ($>10^{11}\ M_{\sun}$) is 2:3, where as at masses below $10^{11}\
  M_{\sun}$ this ratio narrows to 1:3, or more flattened galaxies.  In
  an entirely analogous manner we select a high-redshift sample ($0.6
  < z < 0.8$) from two deep-field surveys with multi-wavelength and
  HST/ACS imaging: GEMS and COSMOS.  We find a seemingly universal
  mass of $\sim 10^{11}\ M_{\sun}$ for highly flatted early-type
  systems at all redshifts.  This implies that the process that grows
  an early-type galaxy above this ceiling mass, irrespective of cosmic
  epoch, involves forming round systems.  Using both parametric and
  non-parametric tests, we find no evolution in the projected axis-ratio
  distribution for galaxies with masses $>3\times 10^{10}\ M_{\sun}$
  with redshift.  At the same time, our samples imply an increase of
  2-3$\times$ in co-moving number density for early-type galaxies at
  masses $>3\times 10^{10}\ M_{\sun}$, in agreement with previous
  studies.  Given the direct connection between the axis-ratio
  distribution and the underlying bulge-to-disk ratio distribution,
  our findings imply that the number density evolution of early-type
  galaxies is not exclusively driven by the emergence of either bulge-
  or disk-dominated galaxies, but rather by a balanced mix that
  depends only on the stellar mass of the galaxy. The challenge for
  galaxy formation models is to reproduce this overall non-evolving
  ratio of flattened to round early-type galaxies in the context of a
  continually growing population.
\end{abstract}

\section{Introduction}

At low redshifts, by number the dominant early-type galaxy is a disky
system, commonly classified as S0 galaxies
\citep{dressler1980a,marinoni1999} or disky elliptical galaxies
\citep[see][for a summary]{kormendy1989}.  These galaxies are smooth,
but have significant rotational support
\citep{krajnovic2008,emsellem2011}. In addition, these galaxies have bulges, which
can often contain 50\% of the light \citep[e.g.][]{laurikainen2010}, but
the presence of disk sets these galaxies apart from the more massive
elliptical galaxies.  In contrast, the most massive galaxies are
generally much rounder systems that are triaxial
\citep[e.g.,][]{franx1991,jf94,vincent2005,vanderwel2009b,bernardi2010,emsellem2011}.
The formation process of the early-type population must not only stop
star-formation, but must form galaxies with a variety of apparent
shapes.

Recently, \citet[][vdW09]{vanderwel2009b} found that there is a
threshold mass for the formation of early-type disk galaxies, a result
which has been indicated in earlier work \citep{jf94}.
For galaxies with stellar masses below $\sim 10^{11}\ M_{\sun}$, there
is a broad distribution of projected axis-ratios.  This implies that
intrinsically flat systems, such disks or flattened
ellipticals, populate these masses.  Above that mass, however,
galaxies become distinctly rounder.  \citet{bernardi2010} find that
this threshold mass is apparent in not just the projected axis-ratio, but in
properties of the stellar population such as the color and
color-gradients.  The implication is that the most massive passively
evolving galaxies are the result of a different formation process than
the galaxies below this mass threshold.  If one assumes that round
galaxies are a result of mergers, than the apparent ceiling in the
mass distribution of disk galaxies reflects the limit in mass for
disky systems.  The process of galaxy formation, therefore, sets a
mass scale above which multiple mergers are the apparently only method
of mass assembly.

There is significant evidence for evolution of number density of
passively evolving $L^{\star}$ galaxies in field surveys
\citep{wolf2003,bell2004,borch2006,brown2007,faber2007,cirasuolo2007,ilbert2010,brammer2011}.
This evolution can occur via two paths, namely the process of merging
building up red-sequence or of star-forming spiral galaxies being
transformed into red-sequence galaxies with the cessation of
star-formation \citep[see][for a more thorough discussion]{faber2007}.
By examining the evolution of the threshold mass as found by vdW09, we
can trace out the relative importance of these different formation
channels over time.  We have investigated this question by compiling
mass-limited samples of passively evolving galaxies from $z=0$ to
$z=1$ in a variety of environments.  We will use the projected axis-ratio
measurements, which have been robustly tested in previous work
\citep[][hereafter H09]{holden2007,holden2009}, to determine the
evolution in the distribution of disky and apparently round galaxies
with redshift.

Throughout this paper, we assume $\Omega_m = 0.27$, $\Omega_{\Lambda}
= 0.73$ and $H_o = 71\ {\rm km\ s^{-1}\ Mpc^{-1}}$.  All stellar mass
estimates are done using a Chabrier initial mass function
\citep[IMF,][]{chabrier2003}.

\section{Data}
\label{data}

\subsection{Parent Sample Selection}

We construct stellar-mass limited samples of early-type or quiescent
field galaxies with HST imaging at redshifts $z\sim 0.7$ from GEMS
\citep{rix2004}, COSMOS \citep{scoville2007a}, and complement this
with a sample of low-redshift counterparts from the SDSS
\citep{york2000}.  Before we estimate stellar masses and select
quiescent galaxies, we construct parent samples from existing
catalogs. From the SDSS DR7 \citep{dr7} we select all
galaxies\footnote{We use the 'Galaxy' table of the DR7 data release
  accessible through CAS jobs.} in the redshift range $0.04 < z
<0.08$.

For GEMS, which overlaps largely with the extended Chandra Deep
Field-South (E-CDFS), we take the public catalog with photometry and
derived quantities such as photometric redshifts from
\citet{cardamone2010}.  As a pre-selection we require that galaxies
are in the photometric redshift range $0.6 < z < 0.8$, have 2$\sigma$
K-band detections, which removes those objects in noisy parts of the
image, lie within the B, V, and R images, and have good photometric
redshifts. ($\chi^2 < 250$, which removes objects with deviating
spectral energy distributions such as AGN).  

For COSMOS, we use the public photometric redshift catalog by
\citet{ilbert2009}.  Again, as a pre-selection we require galaxies to
lie within the photometric redshift range $0.6 < z < 0.8$, have errors
on the r, i, z, J, and K magnitudes less than 0.3 mag, and errors on
the g magnitude less than 0.5 mag.

\subsection{Rest-frame Magnitudes}

For each galaxy, we need to convert the observed photometry into the
equivalent photometry as if the galaxy were observed at a redshift of
$z=0$.  We will refer to these magnitudes and colors with subscript 0,
such as $(u-r)_0$.  To compute these values, we build on the approach
we have used in the past, see for example \citet{blakeslee2005}, or
\citet{holden2010}.  To compute the conversion, we use \citet{bc03}
stellar populations with a variety of $\tau$ models and a range of
metallicities.  At each redshift of interest, we compute the relation
between a pair of observed magnitudes and a rest-frame magnitude of
the following form \( m_{0} = m_{obs_1} + A (m_{obs_1} - m_{obs_2}) +
B \) where $m_{0}$ is the magnitude of interest at $z=0$ and
$m_{obs_1}$ is the observed magnitude that is closest to covering the
same portion of the galaxy spectral energy distribution at a given
redshift $z$.  We then fit the distribution of coefficients, $A$ and
$B$, as splines as a function of redshift $z$ and observed color
$m_{obs_1} - m_{obs_2}$ .  By fitting a spline to the coefficients, we
can calculate a unique conversion for each galaxy based on its
observed redshift and colors.  Then, we add $2.5 \log_{10} (1+z)$ to
each magnitude.

\subsection{Color-Color Selection of Quiescent Galaxies}

To select galaxies that have very little or no star-formation
activity, we follow the now commonly adopted approach to define a
region in color-color space that effectively separates such galaxies
from star-forming galaxies \citep{wuyts2007, williams2009, wolf2009,
  whitaker2010, patel2010}.  In Figure \ref{urz} we show the
rest-frame $(u-r)_0$ and $(r-z)_0$ color distribution of the full SDSS
spectroscopic galaxy sample at $0.04 < z < 0.08$, computed as
described above from the SDSS model magnitudes.  Quiescent galaxies
populate a small region of color-color space as indicated by the
pronounced dark area in the figure. Star-forming galaxies populate a
much more extended, yet fairly narrow sequence.  \citet{patel2011} has
shown that the galaxies in the quiescent region are morphologically
like early-type or elliptical and S0 galaxies in the redshift range of
$0.6 <z < 0.8$.  Therefore, for the rest of this paper, we will refer
to galaxies in that region as early-type systems.

\subsubsection{Determining the Color-color Selection Criteria}

\begin{figure*}
\begin{center}
\includegraphics[width=6in]{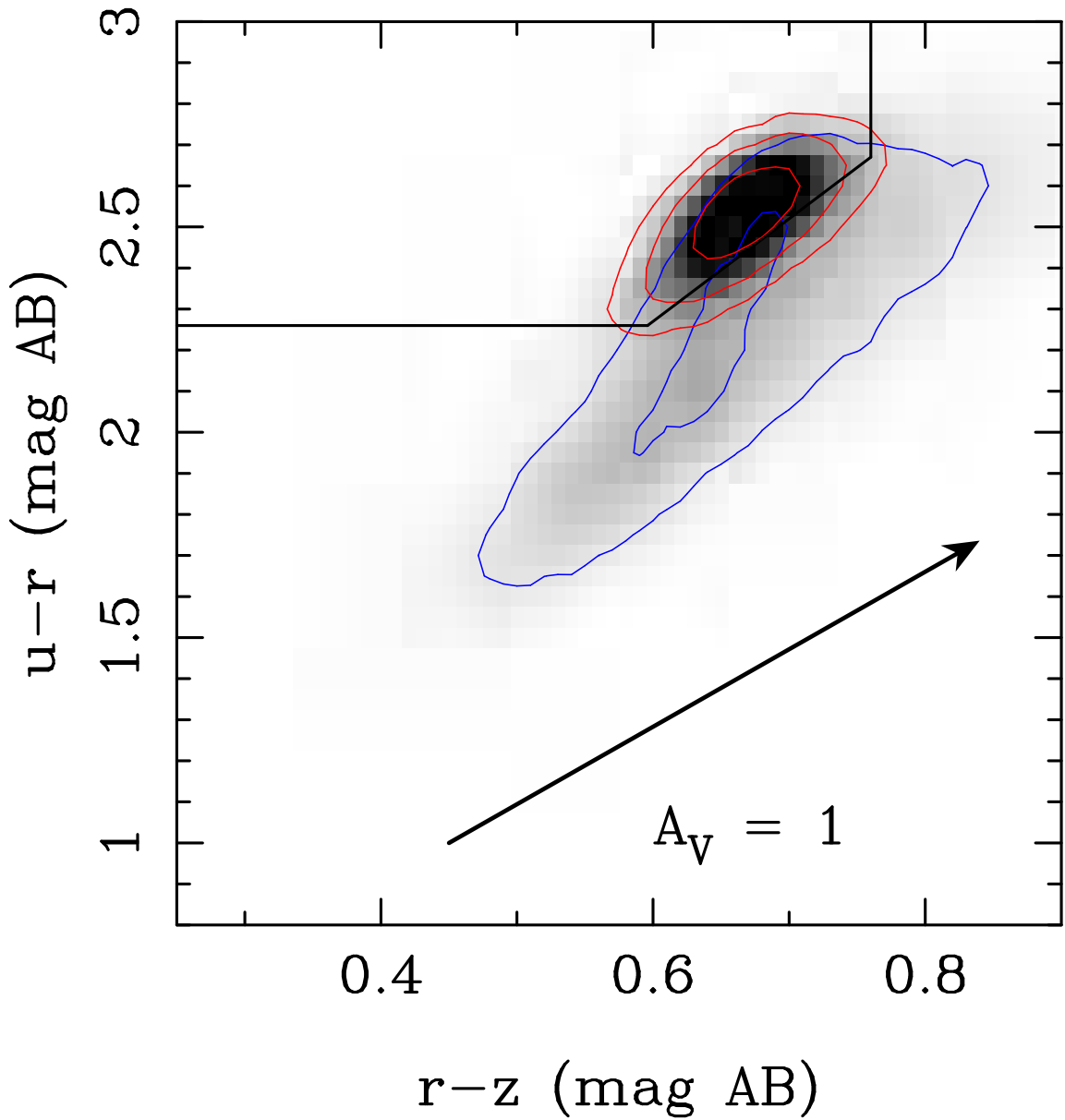}
\end{center}
\caption{The $u-r$ versus $r-z$ relation for galaxies in the $0.04 < z
  < 0.08$ from our SDSS sample in grey scale.  Contours show the
  density of galaxies with \ha\ (red) and without \ha\ (blue).  The
  contours are at 100, 300, and 1000 galaxies per color bin.  The
  black lines have been chosen to minimize the difference between the
  number of galaxies with \ha\ within the box and the number of
  galaxies without \ha\ outside the box. This ensures robust
  separation of star-forming, late-type galaxies and quiescent,
  early-type galaxies even when spectroscopic information is not
  available, for example at higher redshifts.  \label{urz}}
\end{figure*}

In order to define the optimal color-color selection criteria and
establish the reliability of using these criteria to select quiescent
galaxies, we compare the color-color distribution with the H$\alpha$
detection rate.  In Figure \ref{urz} we show the color-color
distribution of galaxies that are detected in H$\alpha$ (at the
2.5$\sigma$ level: $EW(H\alpha) > 2.5 \sigma_{EW(H\alpha)}$) with blue
contours and those that are below that detection threshold in red.
%This is similar to but different than that used in
%\citet{graves2009a,graves2009b}. - fix

We define a polygon to photometrically select quiescent galaxies as
shown in the figure, with the location of the three segments -- but
not the slope of the tilted segment, which is chosen to run parallel
to the star-forming sequence -- as free parameters.  We then calculate
the fraction of H$\alpha$ emitters contained within the polygon (the
contaminating fraction $f_c$) and the fraction of H$\alpha$-less
galaxies outside the polygon (the missing fraction $f_m$) for a each
set of polygon parameters.  The optimal parameter values found by
minimizing $f_c+f_m+abs(f_c-f_m)$, which ensures that $f_c$ and $f_m$
are essentially equal while their sum is minimized.

For the full spectroscopic sample, the optimal color-color selection
criteria are described by the polygon $(u-r)_0 > 2.26$, $(r-z)_0 <
0.75$, $(u-r)_0 > 0.76 + 2.5 (r-z)_0$, with contaminating and missing
fractions of $f_c \sim f_m \sim 0.18$.  The color-color distribution
peaks at $(u-r)_0=2.55$ and $(r-z)_0=0.67$, which is computed by
finding the maximum in density of the color-color distribution after
integrating over a circle with radius 0.03 mag, the approximate
relative error in the colors.  If we pre-select galaxies with a
certain minimum stellar mass (see below), the polygon does not change
by more than 0.02 mag up until the minimum mass reaches very large
values ($\sim 10^{11}~M_{\sun}$).  The contaminating and missing
fractions have a mild dependence, and increase from $\sim 0.18$ to
$\sim 0.20$ if the minimum mass increases from $2\times
10^{10}~M_{\sun}$ to $10^{11}~M_{\sun}$.

A simple red-sequence selection has a much higher contamination
rate.  If we use only the $u-r$ selection, our sample has a higher
success rate, we miss only 0.03 of the early-type sample, but
at a cost of a contamination fraction of 0.37.  Such a high fraction
of star-forming galaxies in red-sequence selections has been seen
before \citep[e.g.,][]{maller2009}.

\subsubsection{The Color-color selection for $0.6 < z< 0.8$ Sample}

\begin{figure*}
\begin{center}
\includegraphics[width=6.4in]{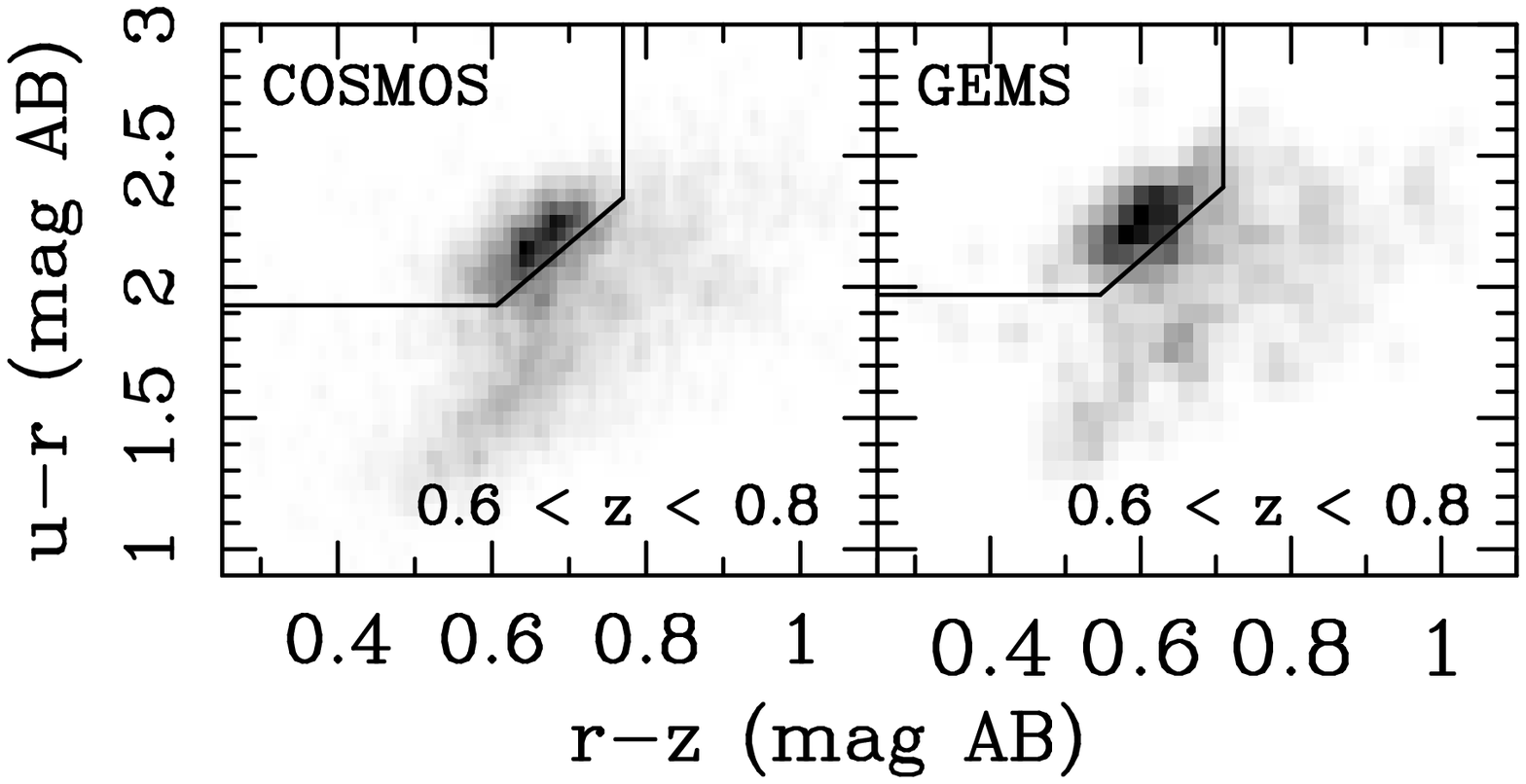}
\end{center}
\caption{The color-color relation for galaxies in the $0.6 < z <
  0.8$ for COSMOS (left) and GEMS (right). The black lines show our
  color selection based on the lines from Fig.\ \ref{urz}, shifted
  by the median colors of the quiescent population.  As with Figure \ref{urz}, the data have
  been lightly smoothed  with a 1.5 pixel FWHM Gaussian before we plot the grey scales.
  \label{highz}}
\end{figure*}

We use the above results for the SDSS sample to define the appropriate
color-color selection criteria for the GEMS and COSMOS surveys.  For
consistency with the SDSS, we derive rest-frame $(u-r)_0$ and
$(r-z)_0$ color (as well as stellar masses, see below) using the
public photometric redshift estimates for both the GEMS and COSMOS
surveys. The color-color distributions for the GEMS and COSMOS samples
are shown in Figure \ref{highz}, where a similar peaked distribution
in color-color space indicates the presence of a quiescent population
of galaxies.  Of course, we do not have the luxury to compare
color-color selection criteria with H$\alpha$ line strengths at $z\sim0.7$.  Instead
we shift the polygon defined above in both $(u-r)_0$ and $(r-z)_0$ by
the difference in the location of the density peaks in the color-color
distribution between high-$z$ samples and the SDSS sample. For GEMS,
the color-color distribution peaks at $(u-r)_0=2.24$ and
$(r-z)_0=0.61$, and for COSMOS at $(u-r)_0=2.28$ and $(r-z)_0=0.70$.
As with the SDSS, we determine these centroids by integrating over a
circle with radius 0.03 mag.  These numbers do not change by more than
0.01-0.02 mag in case errors in the color of 0.05 ma are assumed.

There is an unfortunate difference in the locations of the quiescent
galaxies in color-color space between the GEMS and COSMOS samples.
This difference should be attributed to systematic effects in the
photometry, which would be of the level of $\sim 10\%$.  However, our
method to define the color-color selection criteria for quiescent
galaxies ensures that these systematic effects do not affect our
sample selection and analysis.

\subsection{Stellar Mass Estimates}
\label{stellarmass}

The DR7 catalog of \citet{brinchmann2004} provides stellar mass
estimates for the galaxies in our SDSS sample.  These estimates use a
new methodology for DR7, namely the spectral energy distributions are
fit to the broad band photometry similar to the implementation of
\citet{salim2007} instead of using only the spectra as was done in the
past.  These differences are documented at
http://www.mpa-garching.mpg.de/SDSS/DR7/.  The model fit to the
photometric data yields a distribution.  We select the median mass value as
the best estimate with the 16 and 84 percentile confidence limits as
estimates of the errors.

\begin{figure}
\begin{center}
\includegraphics[width=3in]{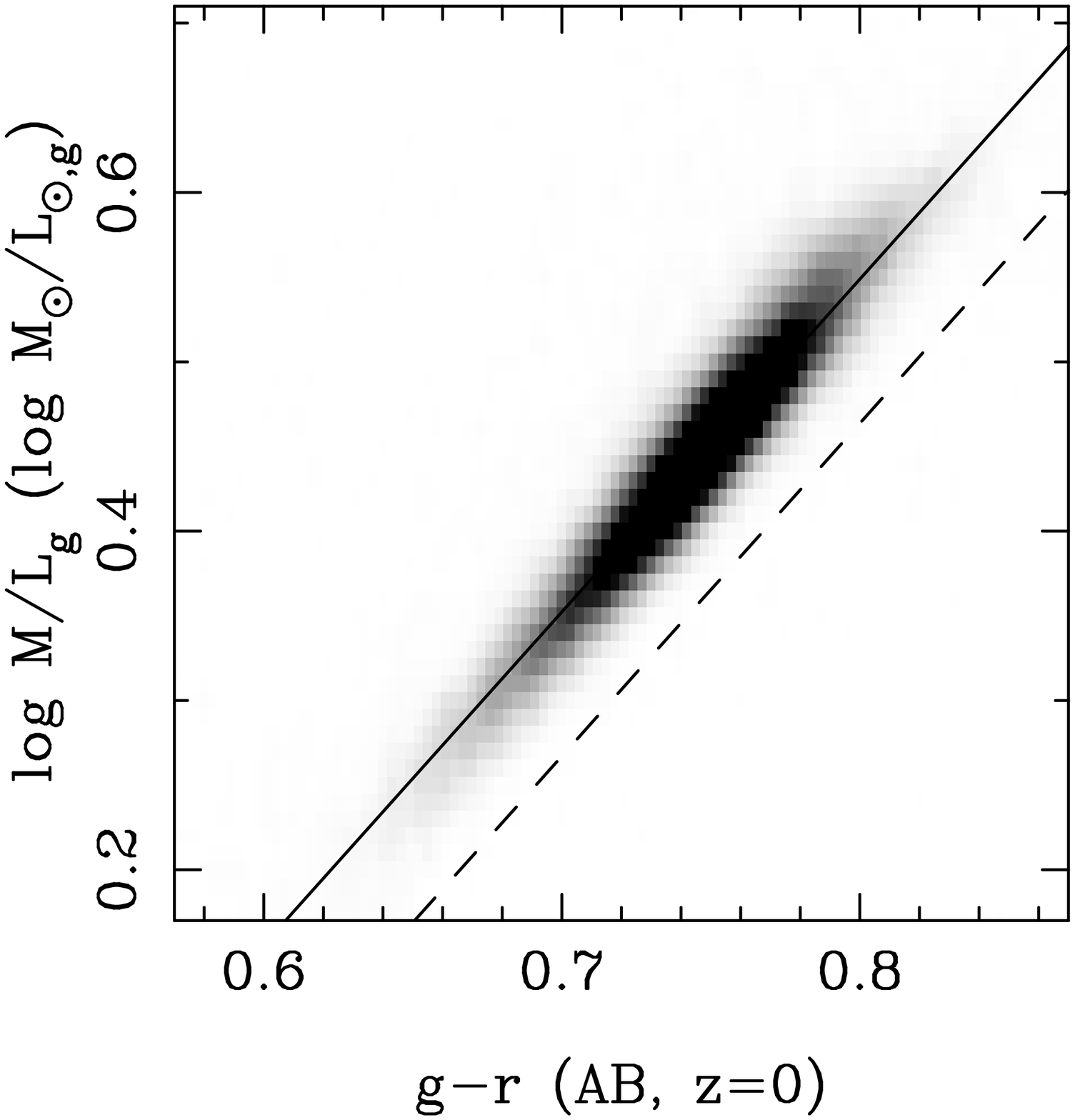}
\end{center}
\caption{The $g-r$-$M/L_g$ relation for early-type galaxies in SDSS
  ($M>1.25\times10^{10}\ M_{\sun}$; $0.04 < z < 0.08$) selected by
  their $u-r$ and $r-z$ colors (see Figure \ref{urz}). The solid line is the
  best-fitting relation. To estimate masses for our $z\sim0.7$ masses
  from COSMOS and GEMS in a consistent manner we use the dashed line
  which has the same slope but a different zero point to compensate
  for evolution in the color and the mass-to-light ratio as explained
  in the text.  \label{mlrgmr}}

\end{figure}

In Figure \ref{mlrgmr} can be seen that quiescent galaxies obey a
tight relation between color and stellar mass-to-light ratio -- this
is obviously inherent to the method.  We derive a linear relation by
computing the biweight mean in narrow color bins (0.02 mag wide) over
the range $0.64 < (g-r)_0 < 0.82$, where the sequence is well
populated.  A fit that minimizes the scatter in the biweight yields:
$\log_{10} (M/L_g) = -1.024 + 1.966 (g-r)_0$.  The scatter decreases from
$\sigma=0.10$ to $\sigma = 0.04$ in $\log_{10} (M/L_g)$ over the probed color
range.

For our high redshift samples, we would like to estimate stellar
masses consistently.  However, we cannot directly use the above
relation because quiescent galaxies at $z\sim 0.7$ have had different
star formation histories than their present-day counterparts, because,
if nothing else, the universe was younger.  We use independent
constraints on the evolution of $(g-r)_0$ and $M/L_g$ to convert the
above relation between color and $M/L$ for $z\sim 0.06$ galaxies to
one appropriate for $z\sim 0.7$ galaxies.  We take the evolution in
$M/L_g$ from the recently derived fundamental plane evolution by
\citet{holden2010}: $d \log_{10} M/L_g = (-0.60 \pm 0.04) z$.  We infer
the $(g-r)_0$ color evolution by computing the biweight mean of the
photometrically selected SDSS, GEMS, and COSMOS samples, for which
find, respectively, $(g-r)_0=0.75$, 0.61, and 0.60.  Thus, we shift
the relation between $(g-r)_0$ and $M/L_g$ shown in Figure
\ref{mlrgmr} to the left by 0.14 (COSMOS) or 0.15 (GEMS) mag, and down
by 0.38 dex, resulting in stellar mass estimates that are $\sim 0.1$
dex lower at $z\sim 0.7$ than those for $z\sim 0.06$ at the same
$(g-r)_0$ color.

To estimate errors on the stellar masses for the high redshift
samples, we use the scatter in the $M/L_g$ versus $g-r$ relation at a
the color of a given galaxy.  This scatter is done in the relation
after it is shifted by the mean $g-r$ color and $M/L$ evolution.  We
then add in quadrature the uncertainties in the zero point of the
$M/L$ evolution.  Thus the error in the high redshift stellar masses
range from 0.06 to 0.11 dex depending on the color of the galaxy.  

% This small difference in the stellar masses we find when using the
% dynamically measured $M/L$ evolution as compared to assuming the
% $z\sim0$ relation between $g-r$ and $M/L_g$ would imply little or no
% mass growth for individual quiescent galaxies between $z\simeq 0.7$
% and today, a not surprising result.  

If use the $M/L$ evolution from the field sample of
\citet{vanderwel2005}, we would estimate smaller masses at $z=0.7$ by
0.02 dex, a small systematic shift.  This difference is in good
agreement with the statistical errors of the \citet{vanderwel2005} and
\citet{holden2010} samples.

\subsection{Projected Axis-Ratio Measurements}

Our projected axis-ratio measurements, $q_{proj}$, come from two approaches.  For the
GEMS and COSMOS samples, these values are the result of fitting
S{\'e}rsic models \citep{sersic} to the two-dimensional images using
the software {\tt GALFIT} \citep{peng2002}.  For the SDSS catalog, we
use the estimates from fitting a de Vaucouleurs model  as part of the
SDSS DR7 photometric pipeline \citep{dr7}.

The GEMS fits are contained in the catalog of \citet{haussler2007}.
These are done using {\tt GALFIT} as part of larger package known as
{\tt
  GALAPAGOS}\footnote{http://astro-staff.uibk.ac.at/$\sim$m.barden/galapagos/}.
{\tt GALAPAGOS} has the advantage of automating the catalog
construction and model fitting process.  The software automatically
fits neighboring objects and incorporates its own sky subtraction
algorithm in order to ensure robust parameter measurements.  A more
thorough discussion can be found in \citet{haussler2007}.

\citet{griffith2010} used {\tt GALAPAGOS} to produce publicly
available catalogs for both COSMOS and GEMS.  We will use the catalog
of \citet{griffith2010} for COSMOS and the catalog of
\citet{haussler2007} for GEMS, though,as we show below, there is no
measurable difference between the \citet{griffith2010} and the
\citet{haussler2007} catalogs.

\subsubsection{Consistency and Reliability of Projected Axis-Ratio Measurements}

Our analysis hinges on the consistent measurement of the axis-ratios
of galaxies at very different redshifts and from very different
imaging data sets.  In \citet{holden2009}, we tested our our
axis-ratio measurements from {\tt GALFIT} with simulations of
observations of high redshift galaxies using real low redshift
galaxies as templates.  We found these measurements to be robust, with
a negligible shift in the axis-ratio from $z=0$ to $z=1$ of $\delta q_{proj}
\simeq -0.01$ with a scatter of $\sigma_q \simeq 0.01 - 0.03$
depending on galaxy magnitude.

%The SDSS provides two measures of the axis-ratio, one based on a de
%Vaucouleurs model fit and one based on an exponential disk model.  We
%compare these values.

In addition to data-related differences, the fitting algorithms also
differ between the low- and high-redshift galaxy samples. We use the
adaption of {\tt GALAPAGOS} for SDSS imaging from \citet{guo2009} to
measure the axis-ratios of a sub-sample of our SDSS galaxies in a
manner that is fully consistent with the treatment of the
high-redshift galaxies.  For small axis-ratios, systematic differences
are expected to be largest. Therefore, we select 412 SDSS galaxies
from our sample with axis-ratios $0.3 < q_{proj} < 0.305$ as determined by
the SDSS pipeline.  These represent the extreme end of the
distribution, where systematic differences in fitting procedures or
the point spread function should be the most manifest.  The axis-ratio
measurements from {\tt GALAPAGOS} are fully consistent with the SDSS
pipeline measurements: $\delta q_{proj} = q_{GALAPAGOS} - q_{SDSS} = 0.01 \pm
0.03$, where 0.03 is the root mean-squared scatter.

{\tt GALAPAGOS} includes many free parameters that affect source
detection, sky subtraction and treatment of neighbors.  However,
regardless of the different set ups that \citet{haussler2007} and
\citet{griffith2010} used for the GEMS data set, there is no
systematic difference between the two catalogs for the objects in our
sample ($\delta q_{proj}= 0.02 \pm 0.03$).

A final test of the robustness of our measurements is a comparison
between axis-ratio measurements from those galaxies in the GEMS that
lie within the much deeper ACS images from GOODS
\citep{giavalisco2004a}.  The difference is $\delta q_{proj} = -0.01$ with a
scatter of $\sigma_q = 0.05$. The negligible systematic difference is
encouraging.

Summarizing, all tests and simulations stress that our axis-ratio
measurements across the different data sets at different redshifts and
performed with different algorithms, are internally fully consistent.

\subsection{Completeness}
\label{complete}

We would like to measure evolution of the axis-ratio as a function of
stellar mass.  In order to probe as far down the mass function as
possible, we need to limit our sample to above the mass limit where
the initial survey photometry is complete. 

For the SDSS, we find that our sample will have no bias as a function
of redshift for early-type galaxies with a mass of $3\times 10^{10}\
M_{\sun}$ for the whole redshift range of $0.04 < z < 0.08$.  This
means that galaxies above that mass range have an equal probability of
being included regardless of redshift.  For the mass of $1.25\times
10^{10}\ M_{\sun}$, this is correct for only the redshift range of
$0.04 < z < 0.06$.  For the rest of this paper, we will include the
whole sample to first mass limit, $3\times 10^{10}\ M_{\sun}$.  For
the mass range $1.25\times 10^{10}\ M_{\sun} < M < 3\times 10^{10}\
M_{\sun}$ we limit the redshift range to $0.04 < z < 0.06$.

At higher redshifts, we have two different samples with two different
selections.  The GEMS sample was selected using a $R$ limiting
magnitude.  In contrast, the catalog from COSMOS we use was based on a
combination of $I$ and 3.6 $\mu$m imaging.  Combining the completeness
computations from \citet{ilbert2010} with the 3.6 $\mu$m magnitude
distribution of galaxies with masses $10^{10} < M/M_{\sun} < 1.5\times
10^{10}$, we infer that our sample is complete in this mass
range.  We adopt $1.25\times 10^{10}\ M_{\sun}$ as our mass limit.
From a similar estimate for the GEMS sample we infer a completeness
limit of $3\times 10^{10}\ M_{\sun}$.  For the rest of this paper,
when we compare samples with masses $>3\times 10^{10}\ M_{\sun}$, we
will be comparing the whole SDSS sample with the combined GEMS and
COSMOS sample.  For masses below that limit, we are only comparing a
subset of the SDSS sample with $0.04 < z < 0.06$ with the COSMOS
sample.

\subsection{Axis-Ratio Dependence of the Observations}

In star-forming disk galaxies, dust is often concentrated in the
mid-plane.  This causes a strong color-dependence in the axis-ratio.
For example, \citet{maller2009} finds a slope of $\sim$-0.5 in the
$u-r$ color with $q_{proj}$.

For our sample of early-type SDSS galaxies, we find that the slopes of
the $u-r$ axis-ratio relation are small, -0.07 to -0.09 magnitudes,
depending on galaxy mass.  Thus, rounder galaxies are mildly bluer
than thinner galaxies. These shallow slopes for our early-type population
has two implications.  First, the completeness our sample is not
drastically impacted by the apparent axis-ratio of the galaxy.
Second, that our sample of early-type galaxies are optically thin.  This
second criteria implies that the selection in color-color space
removes galaxies with significant amounts of gas and dust, which is
typical for passively-evolving systems.

Though the slopes of the $u-r$ color with axis-ratio are small in our
SDSS DR7 sample, they still could be the result of our color-color
selection.  Therefore, we measured the relation of the $u-r$ color
with axis-ratio for the sample of galaxies with \ha\ for comparison.
We found statistically indistinguishable slopes in the $u-r$ versus
$q_{proj}$ relation.  

\section{Intrinsic Shapes of $0.04 < z < 0.08$ Early-Type Galaxies}

\begin{figure*}
 \begin{center}
 \includegraphics[width=6in]{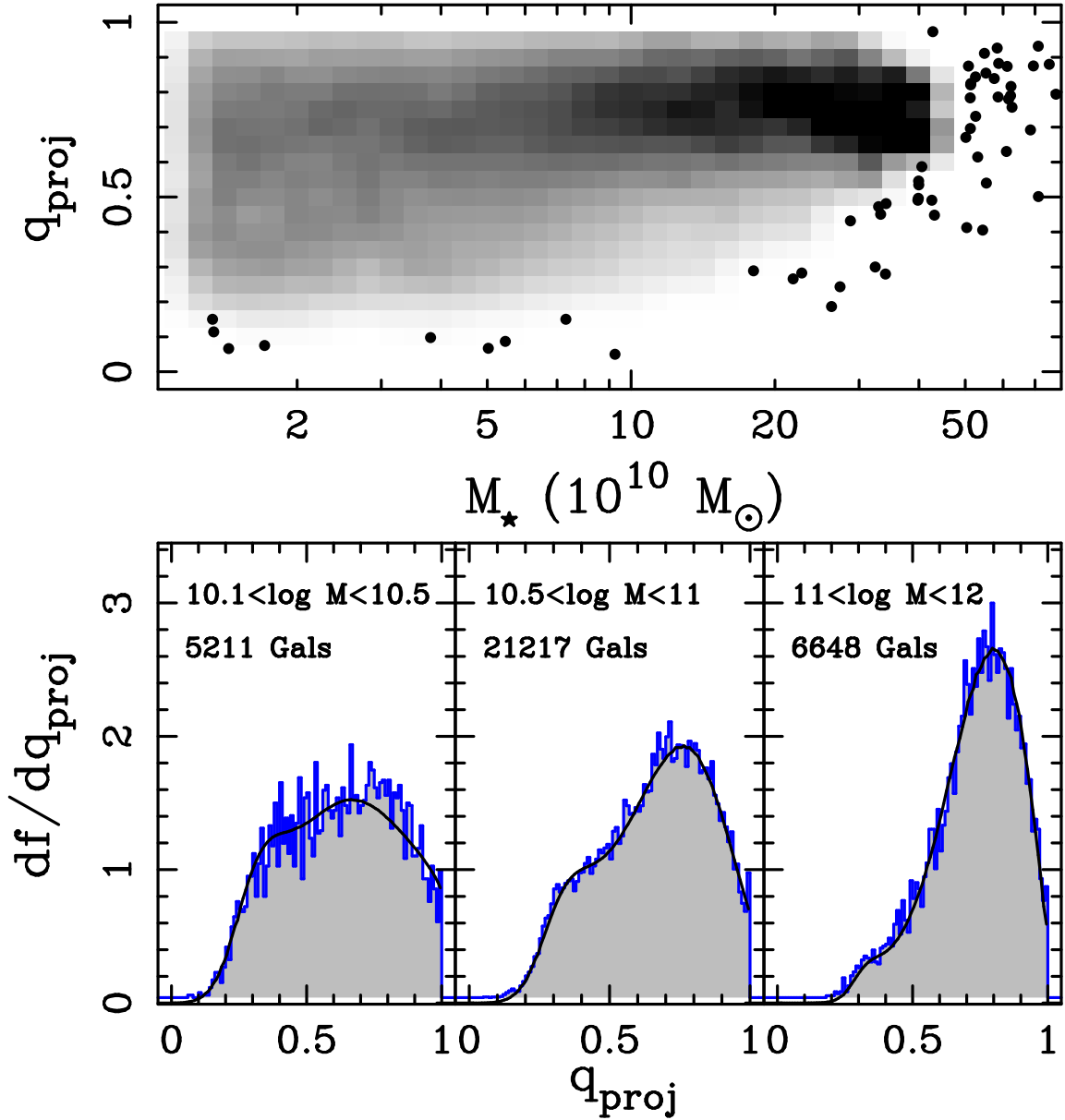}
 \end{center}
 \caption{The distribution of axis-ratios as a function of stellar
   mass for quiescent galaxies in the SDSS (top). When drawing the
   grey scale we normalize by the number of data points in each mass
   bin.  Individual points represent galaxies where the number per bin
   would be less than three.  The sample is complete to $3\times
   10^{10}\ M_{\sun}$ for $0.04 < z < 0.08$, and to $1.25\times
   10^{10}\ M_{\sun}$ for $0.04 < z < 0.06$. The narrowing of the
   distribution shows the threshold for early-type, disky galaxies, as
   seen by vdW09 and \citet{bernardi2010}.  In the bottom panels, we
   plot the differential distribution of axis-ratios in three mass
   bins (marked by the vertical lines in the top panel with the log of
   the range in solar mass given). The number of early-type galaxies
  in each bin is given below the mass range. We over plot, in black,
   parametric models of the data described in Sec.\
   \ref{model}. \label{rix}}
 \end{figure*}

We plot the axis-ratio distribution as a function of stellar mass for
the \nsdss\ galaxies in our final SDSS DR7 sample in Figure \ref{rix}. In that
figure, we also plot the axis-ratio distributions in three broad mass
bins. There are two obvious features in Figure \ref{rix}.  At high
stellar mass, there is a clear absence of elongated objects, implying
the absence of high-mass, disk-dominated, passively-evolving galaxies
at both $z=0.6 - 0.8$ and at $z\sim 0$, as was seen in vdW09.  Second,
below a threshold mass of $\sim 10^{11}\ M_{\sun}$, we find a much
broader distribution of galaxy axis-ratios, implying a significant
population of disk-dominated galaxies \citep[][vdW10]{vanderwel2010a}.

\subsection{A Parametric Description of the Data}
\label{model}

We model the observed, projected axis-ratio distribution to infer the
intrinsic axis-ratio distribution, assuming two types of toy models:
triaxial systems and oblate spheroids.  We use triaxial models because
they are well motivated by other results which include galaxy
kinematics \citep[see, for example][]{franx1991,vandenbosch2008}.  We
use the triaxial model of \citet{franx1991} which has two components,
a triaxiality, $T = \frac{a^2 - b^2}{a^2 - c^2}$, and an ellipticity
$\epsilon = 1- c/a$ where $a$, $b$, and $c$ are the three different
axis making up the triaxial system, with $a$ being the largest and $c$
being the smallest.  We assume that both of these quantities are
distribution as Gaussian and so fit $\bar{T}$, $\bar{\epsilon}$,
along with $\sigma_T$ and $\sigma_{\epsilon}$.

Our second model is a subset of the triaxial model, namely the oblate
spheroid model is defined by fewer parameters.  We use the formalism
of \citet{sandage1970} to describe the oblate spheroid component of
the distribution.  The free values are the minimum apparent axis
ratio, $b$, corresponding to thickness of the spheroid.  Once again,
we assume a Gaussian distribution around $\bar{b}$ of size $\sigma_b$.
The oblate spheroid model can be reproduced by a triaxial model with
$T=0$ and $\sigma_T = 0$.

\subsubsection{The Fitting Process}

We fit a number of models to the data simultaneously, ranging from one
to three. If we fit more than one, we add additional fractions, $f$
which is simply the fraction of overall model distribution that is
represented by a given component.  To determine the best fitting
model, we maximize the log likelihood assuming a Poisson distribution.
Because the triaxial models are computationally intensive, we
pre-compute the apparent axis-ratio distribution for a grid of values
of $T$ and $\epsilon$.  To compare our data with models, we must then
bin the data to same binning as our precomputed triaxial models, or
bins of $\delta q_{proj} = 0.01$.  For a given set of model parameters
$M=M(\bar{T_1},\sigma_T,1,\bar{\epsilon_1},\sigma_{\epsilon,1},\bar{T_2},....
)$, we compute the relative probability for each bin in $q_{proj}$.  We
adjust the normalization such that the model has the same number of
galaxies as the input sample, and can now compute the Poisson log
likelihood $\log L_i = n_i \log(m_i) - m_i - \log(n_i!)$ for each bin
$i$ where $n_i$ is the number of galaxies in that bin and $m_i$ is the
number predicted by the model with parameters $M$.  We use a Monte
Carlo Markov Chain to fit the distribution of axis-ratio values.

%As a check on our best fitting models, we bootstrap our
%data and refit the models 100 times.  These bootstrap re-samplings
%recover our best fitting model parameters.  In the tables below, we
%report the mean of the bootstrap results and we use the standard
%deviation of the parameters as our estimates of the uncertainties.

\subsubsection{Fitting the  $z\sim0.06$ Galaxy Sample}

We begin the fitting process with a single model, namely a triaxial
distribution.  In general, when fitting triaxial models to only the
axis-ratio distribution, the triaxiality is not well constrained.
This is a simple consequence of the fact the data are a single,
projected value while our model contains two different axes.  For
galaxies with $M<10^{11}\ M_{\sun}$, we find that a single component
is not an adequate description of the data, with the model producing a
deficit of galaxies at larger $q_{proj}$ values and an overabundance
of galaxies at small $q_{proj}$'s.  Above that mass threshold, a
single triaxial model appears to match the data well, however.

The addition of a second, oblate spheroid model matches the data much
more closely. We plot these model fits, along with the SDSS
data in blue, in Figure \ref{rix}.  Comparing the log likelihood
values of the single model and two component model shows that the two
component model is a better description of the data at a statistically
significant level ($>3\sigma$).  

Finally, we add a third triaxial component.  In general, whether we
fit three independent triaxial components or an oblate disk with two
independent triaxial components, the fitting process de-weights the
third component.  The resulting model is dominated by only two
components, which make up $>$90\% of the model.
For the rest of the paper, we will use the two component model with
one constrained to be an oblate spheroid for simplicity.  

\subsection{Model Results for $z\sim0.06$}

For the highest mass bin, we find a low oblate spheroid fraction, 13
$\pm$ 4 \%.  The best fitting model has $T\simeq0.4$, near a triaxial
model in the formalism of \citet{franx1991}.  This is close to the
results found in that paper, and what we expected for large,
dispersion supported systems.

Below $10^{11}\ M_{\sun}$, we can see the long, flat distribution to
small axis-ratios well described by an oblate spheroid.  In the middle
mass bin, the fraction of oblate spheroids is \dfrac\ and it grows to
\lfrac\ in the lowest mass bin.  Interestingly, for the middle mass
bin, the triaxial component has a very similar
shape distribution as the triaxial component at higher masses.

% \begin{deluxetable*}{lrrrrrrr|r}
% \tablecaption{Parameters for Best-fitting Combination of Oblate
%   Spheroid Disk and Triaxial Bulge Components to the $b/a$
%   Distribution.}
% \tablecolumns{8}
% \tablehead{
% \colhead{Mass Range} & \colhead{$f_{\rm ob}$}\tablenotemark{a}  &  \colhead{$\bar{T}$} &
% \colhead{$\sigma_T$}  & \colhead{$\bar{\epsilon}$} &
% \colhead{$\sigma_{\epsilon}$} & \colhead{$\bar{b}$} & \colhead{$\sigma_b$} 
% & \colhead{$f_{\rm ob}$}\tablenotemark{a,b} \\ 
% \colhead{($M_{\sun}$)} & \colhead{(\%)}  &  \colhead{} &
% \colhead{}  & \colhead{} &\colhead{} & \colhead{} & \colhead{} & \colhead{} \\ 
% }
% \startdata
% $10^{11} < M < 10^{12}$ & 13 $\pm$  1 & 0.45 $\pm$ 0.05 & 0.07
% $\pm$ 0.04  & 0.37 $\pm$ 0.02  & 0.15 $\pm$ 0.01 & 0.29 $\pm$ 0.01 &
%  0.05 $\pm$ 0.02  &  11 $\pm$  4 \\
% $3\times 10^{10} < M < 10^{11}$ & 46 $\pm$ 2 & 0.50 $\pm$ 0.04 & 0.02
% $\pm$ 0.04 & 0.44 $\pm$ 0.03 & 0.16 $\pm$ 0.03 & 0.27 $\pm$ 0.03 &
% 0.07 $\pm$ 0.02 & 47 $\pm$ 10 \\
% $1.25\times 10^{11} < M < 3\times 10^{10}$ & 73 $\pm$ 5 & 0.75 $\pm$
% 0.08 & 0.02 $\pm$ 0.01 & 0.51 $\pm$ 0.04 & 0.14 $\pm$ 0.03 & 0.26 $\pm$
% 0.02  & 0.09 $\pm$ 0.01 & 47 $\pm$ 13 \\
% \enddata
% \tablenotetext{a}{The percentage of the $b/a$ distribution model by a
%   oblate spheroid disk component, which is controlled by the
%   parameters $\bar{b}$ and $\sigma_b$.}
% \tablenotetext{b}{For the $0.6 < z < 0.8$ sample.}
% \label{res}
% \end{deluxetable*}

\begin{figure*}
 \begin{center}
 \includegraphics[width=6in]{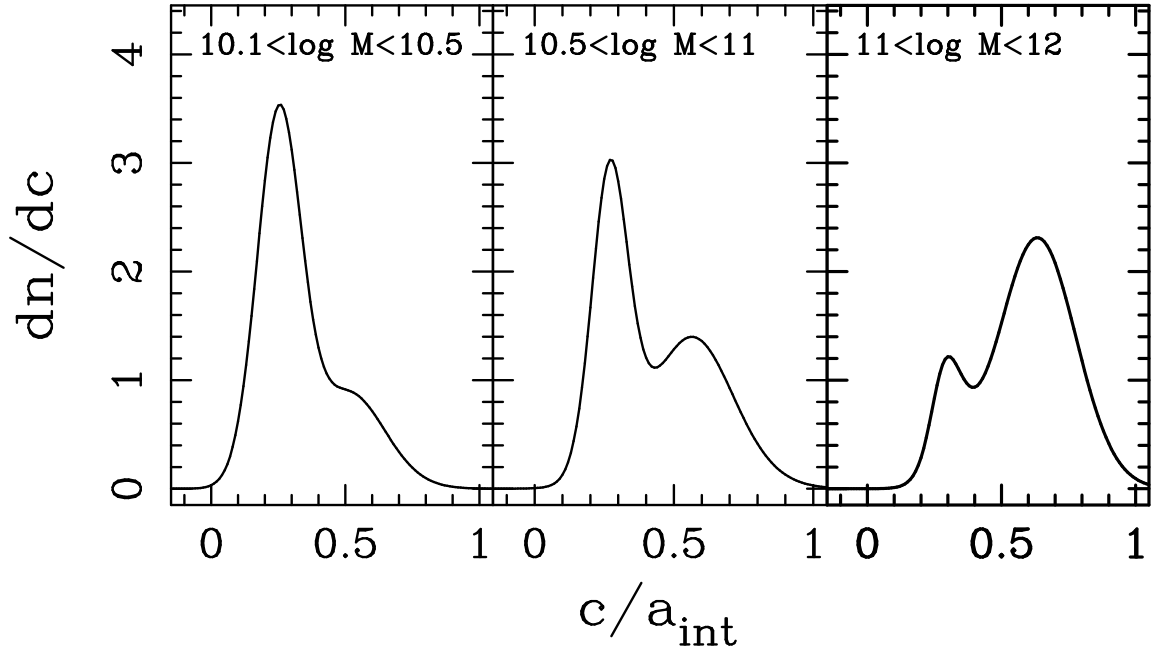}
 \end{center}
 \caption{The intrinsic axis-ratio distribution ($dn/dc$), where $c$
   denotes the smallest-to-largest axis-ratio, in three mass bins for
   early-type galaxies in the SDSS.  These plots represent our
   inferred results weighted by the likelihood distribution combined
   with our Gaussian model parameters.  The model used to infer these
   distributions consist of an oblate component (where $c=b$) and a
   triaxial component (where $c=1-\epsilon$) -- see Sec.\ \ref{model}
   for a detailed description of this model and how the results are
   represented here.  While details depend on the modeling
   assumptions, high-mass early-type galaxies typically have intrinsic
   short-to-long axis-ratios of 2:3, while $M^{\star}$ early-type
   galaxies typically have 1:3.  \label{ctoa}}
 \end{figure*}

One of the more robust parameters for even the triaxial models are the
minor to major axis-ratios.  In Figure \ref{ctoa}, we plot the inferred
distribution from the Monte Carlo Markov Chain of $c$, or the minor to
major axis-ratios.  The two different components to our model are
readily apparent, along with their relative weight.  

Most striking, in even the high mass bin, there are no really round
galaxies.  The modal value for $c$ is inferred to be $c\simeq 0.65$ or
a ratio of $1.5$ to 1.  Such an apparently small minor axis-ratio is
actually in good agreement with the data.  The intermediate axis,
assuming a $T=0.4$ or so, will be $b=0.9$ and so such systems will
often appear to close to but not perfectly round, exactly as we see in
Figure \ref{rix}.  

The second result is that, in all mass bins, there are no very thin
galaxies.  Among the star-forming galaxy population, disks can can
thin $c\simeq 0.2$ , especially at lower masses and even in red
pass-bands \citep{ryden2006,padilla2008}.  Our modal value for the
lowest mass bin in Figure \ref{ctoa} is 0.25 and the median value for
the whole of the distribution is 0.29.  Thus, the low-mass, passively
evolving population is $\sim$50\% thicker than similar mass active
star-forming galaxies which have values more like $0.2$.

\section{Shape Evolution of Early-type Galaxies}

The final $0.6< z < 0.8$ samples contain \nhiz\ galaxies in total
(\ngems\ from GEMS; \ncosmos\ from COSMOS) with masses greater than
$1.25\times 10^{10}\ M_{\sun}$, though see Sec.\ \ref{complete} for
details on the completeness with mass.  We will now use these samples
to measure the evolution of the axis-ratio distributions for samples
in a fixed mass range.

 \begin{figure*}
 \begin{center}
 \includegraphics[width=6in]{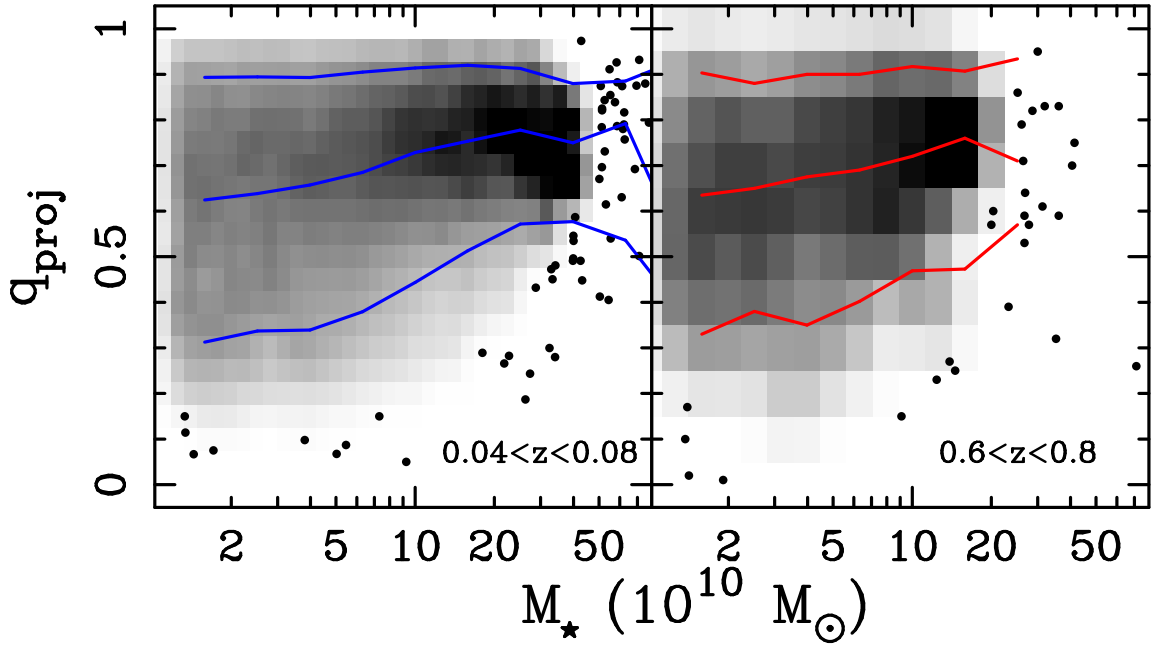}
 \end{center}
 \caption{The distribution of axis-ratios as a function of stellar
   mass for quiescent galaxies for the SDSS (left) and our combined
   COSMOS and GEMS $0.6<z<0.8$ sample (right). The SDSS data are the
   same as shown in Figure \ref{rix}. In each, we over-plot the 10th
   percentile, median and 90th percentile axis-ratios as a function of
   mass for both the SDSS (blue) and the $0.6 < z < 0.8$ sample (red).
   At masses of $>3\times 10^{10}\ M_{\sun}$, the patterns appear
   indistinguishable in the two samples, with a steady tapering above
   $2\times 10^{11}\ M_{\sun}$ yielding an effective ceiling mass for
   high elongated system.  This implies that the threshold for
   early-type, disky galaxies seen at low redshift by vdW09 and
   \citet{bernardi2010} does not significantly evolve out
   $z\simeq0.7$.  Correspondingly, below that ceiling mass, both
   samples show similar distributions.  This implies that at
   $z\sim0.7$ the $M^{\star}$ early-type galaxy is moderately
   ``disky'', as we show in Figure \ref{ctoa}. In contrast, the mass
   density of early-type galaxies grows between $z\simeq 0.7$ and
   $z=0$.  Our result shows that the mass growth must roughly
   preserve the distribution of axis-ratios in each mass bin.  \label{rixtwo}}
 \end{figure*}

\subsection{The Shape Distribution of $0.6<z<0.8$ Galaxies}
\label{threshold}

\begin{figure*}[htbp]
\begin{center}
\includegraphics[width=6.4in]{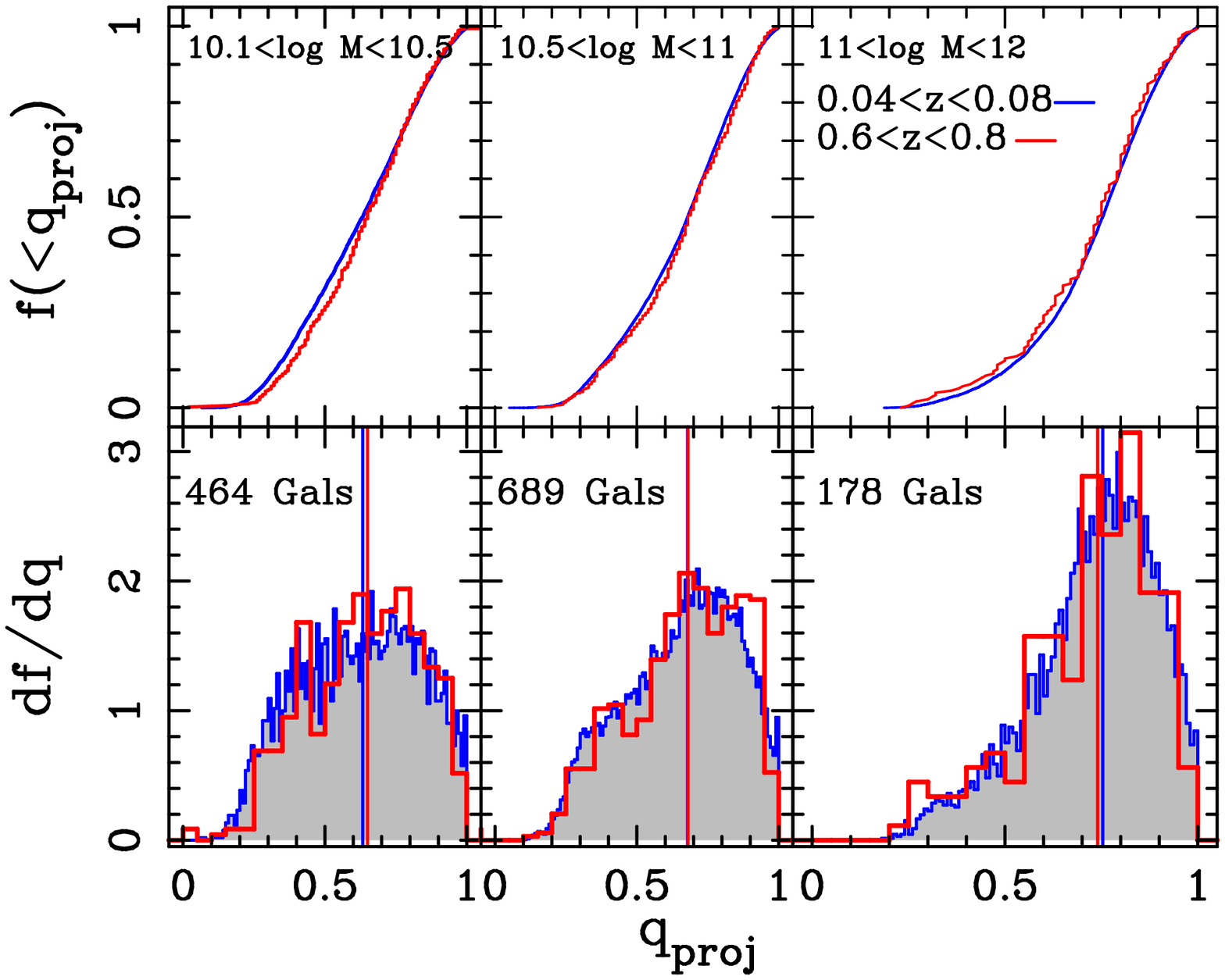}

\end{center}
\caption[f6.ps]{The axis-ratio distribution for our early-type galaxy
  samples in three mass bins, with top and bottom showing the
  normalized cumulative and differential distributions respectively.
  The SDSS data are plotted with a blue line, in both, and grey
  shading in the bottom panels, while, we show our combined sample of
  early-type galaxies at $0.6 < z < 0.8$ in red.  The total number of
  galaxies in the $0.6 < z < 0.8$ sample is given in the bottom
  panels, with the size of the SDSS samples for the same mass range
  given in Figure \ref{rix}.  We plot the median axis-ratio for the
  SDSS with a blue vertical line and the $0.6 < z < 0.8$ sample as a
  red line.  The axis-ratio distribution is statistically similar in
  all three mass bins, with only the lowest mass bin showing a hint of
  evolution.  Because statistically similar models as in Fig.\
  \ref{ctoa} describe the $0.6 < z 0.8$ sample, out to z$\sim$1,
  massive early-types have axis-ratios of 2:3 while early-types around
  $M^{\star}$ show a thinner distribution closer to
  1:3.  \label{dist}}
\end{figure*}

The combined sample of GEMS and COSMOS, along with the corresponding
SDSS sample, can be seen in Figure \ref{rixtwo}.  We plot the axis-ratio
distribution as a function of stellar mass for the $0.04 < z < 0.08$
and $0.6 < z < 0.8$ samples.  In Figure \ref{dist}, we plot axis-ratio
distribution in three broad mass bins, showing both the differential
and cumulative distributions. There are two obvious features in Figure
\ref{rixtwo}.  At high stellar mass, there is a clear absence of
elongated objects, implying the absence of high-mass, disk-dominated,
passively-evolving galaxies at both $z\sim0.7$ and at $z\sim 0$, as
was seen in vdW09.  Second, below a threshold mass of $\sim 10^{11}\
M_{\sun}$, we find a much broader distribution of galaxy axis-ratios,
implying a significant population of disk-dominated galaxies
\citep[][vdW10]{vanderwel2010a}.

We use the parametric models we discuss in Sec.\ \ref{model} to fit
the distribution of data in Fig.\ \ref{dist}.  Because of the much
smaller sample size, we fit the data with both the same models as we
did the SDSS but also freeze some subsets of the parameters.  In every
case, we find that, within the limits of the uncertainties from the
fits, the data can be described by the same model at both redshifts.  

The clear similarity between the axis-ratio distributions of our low-
and high-redshift samples shows that a ceiling mass for quiescent,
disk-dominated galaxies exists at least since $z\sim1$, generalizing
the low-redshift result from vdW09.  It is clear that, above $10^{11}\
M_{\sun}$, we find few flat galaxies.  In contrast, these galaxies
make up a much larger proportion of the population at masses below
$10^{11}\ M_{\sun}$.  Therefore, at $z\sim 0.7$, there is the same
threshold for the population of ``disky'' early-type galaxies, that is found at $z\sim0$.

\subsection{The Axis-Ratio Dependence of the Mass Function of Early-type Galaxies}

Previous work has found a significant amount of evolution in the mass
function of early-type galaxies
\citep{wolf2003,bell2004,borch2006,brown2007,faber2007,cirasuolo2007,ilbert2010,brammer2011}.
Generally, the evolution appears at the lower mass end.  As we have
found that the axis-ratio distribution also changes with mass, with
more disk-like early-types at lower masses, we will investigate if the
mass function evolution is different for different sub-populations as
selected by axis-ratio.

% reread

We fit the mass distribution using a Schechter function.  We use a
standard maximum likelihood approach assuming a Poisson likelihood
model and perform the fits over the mass range where our sample is
volume complete.  Each galaxy has its own error estimate for the mass
measurement, so we convolve the Schechter function individually to
compute the likelihood distribution.  Including the errors in the
fitting process has the advantage of not causing $M^{\star}$ being
forced to higher values because of the occasional statistical
fluctuation in a mass measurement.  

\subsubsection{Edge on systems with $q_{proj}<0.4$}

We select all galaxies with $q_{proj} < 0.4$ and masses $>3\times 10^{10}\
M_{\sun}$.  This selects disk-dominated systems that are viewed close
to edge-on but above the mass limit where we are complete for the
whole volume of both samples.  We fit the mass distribution with a
Schechter function with a fixed value of $\alpha=-0.7$
\citep{bell2003} for both our $0.6 < z < 0.8$ and SDSS sample.  We
find the value of $\log_{10} M^{\star}/M_{\odot} =10.68 \pm 0.10$
(errors come from bootstrapping the data) for quiescent galaxies with
$q_{proj}<0.4$.  This value lies within $1\sigma$ of our $0.04< z < 0.08$
field sample of $\log_{10} M^{\star}/M_{\odot}=10.58\pm 0.01$.  We
confirm the lack of strong evolution by using Monte-Carlo simulations
where we adjust the mass distribution of the $z\simeq0.06$ sample and
create sub-samples of the same size as our high-redshift sample with
the same mass limits.  From this we find that the typical mass of
$q_{proj}<0.4$ galaxies above our mass completeness limit can only shift by
by $\pm 0.06$ dex, $\sim$16\%, in our $0.6 < z < 0.8$ sample.  We also
confirm this result using non-parametric tests, the Kolmogorov-Smirnov
test and the Mann-Whitney test, which also show no significant
difference in the two samples.

\subsubsection{Round systems with $q_{proj}> 0.6$}
\label{roundmf}

We also look for evolution in the apparently round galaxy population,
those with $q_{proj}>0.6$.  From our modeling results in Sec.\ \ref{model},
we expect that this population is a combination of those mostly
triaxial systems, with intrinsic ratios of 2:3, and the more
flattened, or 1:3, population that dominates at lower masses.
Evolution in this population, if not mirrored in the $q_{proj}<0.4$
population, would imply evolution in the more triaxial component of the
population that dominates at high masses.  

When we examine the galaxies that are round, $q_{proj}>0.6$, we find
significant ($>3\sigma$) though mild amount evolution in the mass
function.  For our SDSS sample, we find $\log_{10}
M^{\star}/M_{\odot}=10.93 \pm 0.01$ while in our $0.6 < z < 0.8$
sample we find $\log_{10} M^{\star}/M_{\odot}=10.85 \pm 0.02$.  As
before, we confirm this result at the $>3\sigma$ with the
Kolmogorov-Smirnov test and the Mann-Whitney test.  We also confirm
this result by drawing sub-samples of galaxies from the $q_{proj}>0.6$ and
$0.04< z < 0.08$ SDSS sample of the same size as the $0.6 < z < 0.8$
sample.  We find that these sub-samples recover $\log_{10}
M^{\star}/M_{\odot}=10.93$ with a scatter of $\pm 0.02$ dex.  The
larger question is, does this evolution represent a change in the
galaxy population or is it a result of our measurements?  Our $0.6 < z
< 0.8$ stellar masses have a systematic uncertainty of $\pm 0.04$ dex.
2Thus, the shift in the mass function we find is interesting but not
statistically significant.  Because this an error on the zero point of
stellar masses, which we derive from evolution in the fundamental
plane (see Sec.\ \ref{stellarmass}), the systematic error on the
stellar masses can be lowered in future work.  This will confirm or
refute this apparent evolution shape of the mass function of round
early-type galaxies.

\subsubsection{The whole of the population}

\begin{figure}
\begin{center}
\includegraphics[width=3.4in]{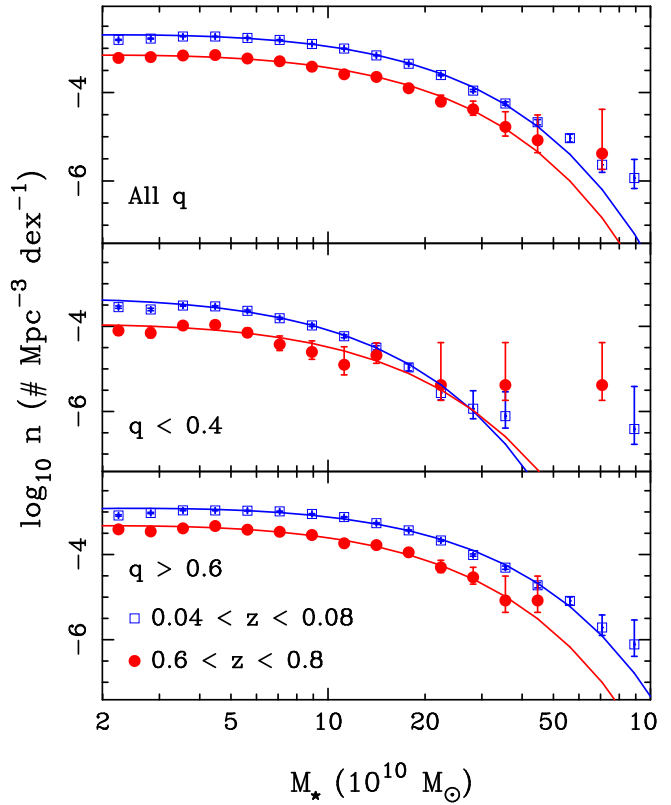}
\end{center}
\caption[]{Mass functions of the $0.04<z<0.08$ (red filled circles)
  and $0.6 < z < 0.8$ (blue open squares) samples, for three different
  axis-ratio ($q_{proj}$ )selections.  We over plot with solid lines
  the best fitting Schechter functions, all assuming a fixed
  $\alpha=-0.7$.  Our fitting process includes the errors on the
  stellar mass estimates.  Reinforcing the visual impression from
  Figure \ref{rixtwo}, the mass function of flat galaxies has a much
  smaller value of $M^{\star}$ than the mass function of almost round
  galaxies.  There is also little evolution in the shape of the mass
  function except for possibly the $q_{proj}>0.6$ sample, see text for
  further discussion.  In general, the volume density of galaxies
 evolves by a factor of $\sim\times$2-3 at $10^{11}\ M_{\sun}$, in good
  agreement with the evolution seen in \citet{bundy2010} or
  \citet{ilbert2010}.  This evolution is the same, regardless of the
  axis-ratio of the population.  This requires that the growth in the
  mass density of galaxies, in the mass range we consider, must occur
  in a manner that preserves the overall shape distribution. }
\label{mf}
\end{figure}

As check on our mass functions, we fit for $M^{\star}$ with a fixed
$\alpha = -0.7$ for whole of our $0.04 < z < 0.08$ sample of passively
evolving galaxies and find $\log_{10} M^{\star}/M_{\sun} = 10.87 \pm
0.01 \ M_{\sun}$, in good agreement with \citet{bell2003} after
accounting for differences in the IMF and $h$.  We find $\log_{10}
M^{\star}/M_{\sun} = 10.85 \pm 0.02 \ M_{\sun}$ for our high redshift
sample, similar to \citet{borch2006}.  In Figure \ref{mf}, we show the
mass functions for three selections in axis-ratio ($q_{proj}<0.4$, $q_{proj}>0.6$
and all galaxies regardless of $q_{proj}$).  We also plot our estimate of the
total number density of galaxies per logarithmic density bin.  It is
clear that we recover the trend in the density evolution of the
passive galaxy population found by \citet{ilbert2010}.  We note that
\citet{ilbert2010} found evolution in $\alpha$.  We find that, because
of our high mass limit of $3\times10^{10}\ M_{\sun}$, we have little
statistical constraint on the best fitting value of $\alpha$.  To
improve our results would require implementing completeness
corrections for both samples.  Nonetheless, we reproduce $M^{\star}$
with the same sample.  Thus, despite our different methodology for
determining stellar masses and different sample definitions, we find
consistent results with other measurements of the mass function.

\subsection{The Bulge-to-Disk Ratio of the Population of Early-type Galaxies}
\label{bdr}

The average axis-ratio of a galaxy population is directly determined
by the populations average bulge-to-disk ratio \citep{binneym1998},
assuming that bulges are drawn from a different axis-ratio
distribution as compared with disks \citep[see Figure 2 of][which shows that this
is true for all but the highest mass galaxies]{dutton2011b}.  Therefore, by examining the evolution of the
axis-ratio of the population, we are determining whether or not the
population becomes more bulge-dominated or disk-dominated as a
function of time, though we cannot determine if this evolution happens
for individual galaxies or because of a changing mix of bulge-to-disk
ratios in the population.

In the range $3\times 10^{10}\ M_{\sun} < M < 10^{11}\ M_{\sun} $, we
find no difference in axis-ratio distribution between the two redshift
slices.  In the lowest-mass bin ($1.25\times 10^{10}\ M_{\sun} < M <
3\times 10^{10}\ M_{\sun} $) we see a small but barely significant
difference between the two samples, suggestive of a more
disk-dominated population in the $0.04< z<0.08$ sample. Because of the
low significance of the difference ($2.4\sigma$) and the fact that it
occurs in the smallest mass bin where the completeness is lowest, we
consider this difference an interesting but tentative result.

\section{Discussion and Conclusions}
\label{disc}

\subsection{A Universal Ceiling Mass for Flattened Early-type Galaxies}

vdW09 observed a ceiling mass of $\sim 10^{11}\ M_{\sun}$ for
disk-dominated, quiescent galaxies in the present-day universe.  In
accordance, \citet{bernardi2010} found that early-type galaxies with
very high masses ($\sim 2\times10^{11}\ M_{\sun}$) differ in many ways
from those with lower masses ($<10^{11}\ M_{\sun}$).  In this paper we
show that a similar transition mass exists at $z\sim 0.7$ and that its
value has not shifted by more than 0.05 dex between $z\sim 0.7$ and
the present.  Thus, at all redshifts, roughly 40\% of the stellar mass
in early-type systems is contained in these relatively round systems.
As expected from Figure \ref{rix}, round systems ($q_{proj} > 0.6$)
have a characteristic mass of $M^{\star} \sim 9 \times 10^{10}\
M_{\sun}$ while highly flattened ($q_{proj}<0.4$), passively evolving
galaxies have $M^{\star} \sim 4 \times 10^{10}\ M_{\sun}$.  We find
no significant evolution in the value for $M^{\star}$ between our two
samples, only evolution in the co-moving number density.

This mass ceiling has the same mass, or, in other words, $M^{\star}$
does not evolve for more elongated or ``disky'' early-type galaxies
despite the growth of the passively evolving population by a factor of
$\sim2-3$ in mass between $z\sim1$ and today \citep[see Figure
\ref{mf}; ][]{wolf2003,bell2004,borch2006,brown2007,faber2007,cirasuolo2007,ilbert2010,brammer2011}.
This has two implications.  First, the progenitors of today's
population of massive early-type population were not more
disk-dominated systems at $z=0.7$ that faded into the
passively-evolving population.  Instead, these galaxies must already
be almost round, roughly 2:3 in intrinsic axis-ratio, galaxies before
the truncation of star-formation \citep[see][for candidate
progenitors]{kocevski2011}.  Second, if merging builds up the
population of galaxies above $\sim 10^{11}\ M_{\sun}$, that merging
most cause them to become rounder systems.

\subsection{Implications for the Formation of Galaxies with $M> 10^{11}\ M_{\sun}$}

At the highest masses, we find that, not only is there a lack of
flattened or ``disky'' galaxies, but that the distribution is
consistent with a largely triaxial population.  This can be seen by
the lack of galaxies that are round in projection galaxies at high
masses in Figure \ref{rix}.  These apparently round galaxies are seen
at lower masses, so we do know that the lower fraction of high mass, round
galaxies not just a systematic measurement error.  \citet{padilla2008} found a
similar result, but the lack of evolution we find means that this
triaxiality is set in the formation of these systems out to $z\sim1$.
This result, when combined with the observed evolution in the
normalization of the mass function, and hints about the shape (see
\ref{roundmf}), this triaxial population is built up over the redshift
range we observe, but is done so in such a way to produce similar
shaped systems over that time.

Massive ellipticals are assumed to form out of multiple mergers of
near equal mass systems and the merger rate is expected to be high
even at redshifts of $z\sim0.7$ \citep[e.g.][]{delucia2007}.  Detailed
simulations with cosmological initial conditions show that additional
mechanisms are required to reproduce the observed shapes and kinematic
profiles of massive ellipticals \citep[e.g.][]{burkert2008,novak2008}.
Minor mergers and tidal encounters also provide a mechanism for making
the most massive quiescent galaxies appear round.
\citet{vulcani2011b} finds that the most massive cluster galaxies,
objects too massive to be in our sample, are less round at high
redshift.  This points to observational evidence of the process of
galaxies becoming rounder with time, possibly because of the
mechanisms suggested in \citet{burkert2008}, but only for the rarest
and most extreme of systems.

\subsection{Evolution of the  $\sim M^{\star}$ Early-type Population}

At masses $<10^{11}\ M_{\sun}$, the early-type population becomes more
and more ``disky''.  This can be seen in two ways, first, we find more
round galaxies, $q_{proj}>0.9$.  Second, we find more flattened systems, $q_{proj}<0.4$.
This can be seen in both the minimum axis-ratio we find in Figure
\ref{rix}, and, the distribution of $c/a$ values we infer from our
parametric modeling in Figure \ref{ctoa}.

Quiescent galaxies with masses that dominate the cosmic stellar mass
budget ($3 \times10^{10}\ M_{\sun}<M_{\star}<10^{11}\ M_{\sun}$) show
a broad but non-evolving range in axis-ratios, at both $z\sim 0.7$ and
$z\sim 0.06$.  The broad range in axis-ratios implies that the
population can form through a number of channels.  Because we find so
little evolution in the axis-ratios, however, means that, whatever the
mechanisms that form early-type galaxies in this mass range, they must
have worked at similar rates across the last 7-8 Gyrs of look-back
time.  This evolution cannot explained entirely by the increase in the
number of bulge-dominated galaxies (say, merger products), nor can it
be explained entirely by the cessation of star formation in
disk-dominated galaxies without structural changes.  Several
evolutionary processes that cause the formation of quiescent galaxies
must contribute in order to explain the unchanging fractions of bulge-
and disk-dominated quiescent galaxies.  Moreover, the relative
importance of the various evolutionary processes has not strongly
changed over the past 7-8 Gyrs.  This is reminiscent of the general
result that the morphological mix of galaxies of these masses does not
significantly change over the same time \citep[vdW07,
H09,][]{bundy2010}.

% How does the cosmological decrease in the gas accretion rate factor
% in here?:

\subsubsection{Growth in the Number Density Growth of Highly Flattened Systems}

Our work finds consistent evolution in the number density of passively
evolving galaxies with redshift.  Most work finds significant
evolution, factors of 2 or 3, in the number density of galaxies in the
mass range of our
sample\citep[e.g.][]{ilbert2010,bundy2010,brammer2011}.  The
combination of a flattened population at low masses with the increased
number density of galaxies with redshift says that, at lower masses,
the buildup of the mass function of passively evolving galaxies, or
early-types, is the build up of passive disk-like galaxies, such as
S0s or ``disky'' ellipticals \citep{bundy2010}.

How can we explain the existence and continued growth of a
population of quiescent, flattened galaxies?  Gas stripping in
group and cluster environments has long been argued to play a role
\citep{spitzer1951}, and was recently shown to explain the existence
of the morphology-density relation \citep{vanderwel2010a}. Our
tentative detection of an increased fraction of ``disky'',
quiescent low-mass galaxies ($<3\times10^{10}\ M_{\sun}$) at late
times may indicate that this process is becoming increasingly
important at late cosmic epochs.

\subsubsection{Structural Properties and Merging}

It is clear that \textit{all} disk-dominated, quiescent
galaxies cannot be the result of gas stripping, especially those outside
massive groups and clusters \citep[e.g.][]{dressler1980a}.  While this
may be feasible in the form of efficient gas stripping from satellite
galaxies, even in sparser group environments \citep{vandenbosch2008},
the observed differences between disky quiescent galaxies and
star-forming spiral galaxies of the same mass imply that the former
are not, generally, stripped versions of the latter.  Quiescent
galaxies typically have fewer bars
\citep{aguerri2009,laurikainen2009}, larger bulges
\citep{dressler1980a,christlein2004,ryden2006,laurikainen2010} and are
more concentrated \citep{bundy2010} than star-forming galaxies, though
many of these properties are mass dependent \citep{cheng2011}.  
Finally, the axis-ratio distributions of star-forming galaxies are
markedly different, much flatter as expected for an oblate spheroid
population alone, then the distributions we observe for early-type
galaxies \citep{ryden2006,padilla2008} Thus,
at least at higher masses, the truncation of star formation must be
intimately linked with bulge growth \citep[e.g.][]{bell2008}, even if
a sizable stellar disks remains intact.

Minor merging may provide a possible path, which would provide a
natural explanation for our observation that the mix of bulge- and
disk-dominated quiescent galaxies remains unchanged at $z\lesssim 1$.
The advantage of this mechanism is that minor merging is common, it
produces most of the growth for massive early-type galaxies \citep[for
example][]{oser2011}.  Second, the amount of minor mergers grow  bulges
\citep[e.g.][]{kauffmann1993,baugh1996}, though that depends on the gas
content of the smaller system \citep{mihos1994b,hopkins2009}.

Because of this, \citet{bundy2010} suggests a two stage scenario.
First, some feedback mechanism causes star-formation to cease.
Second, because lower gas content galaxies have more rapid bulge
growth from minor mergers \citep{hopkins2009}, the resulting remnants
are both passively evolving and bulge-dominated. In fact, bulge growth
through minor merging may cease star formation as a result of gas
exhaustion, some feedback mechanism (possibly AGN), or the
stabilization of a gaseous disk against star formation
\citep[e.g.][]{bower2006,croton2006}.  The main problem with such a
process is, however, that we find no evolution in the overall
axis-ratio distribution.  This means this two stage process must
produce as many bulge-dominated systems as new disk-dominated systems
are added to the early-type galaxy population.  If the rate of these
two processes are not in good agreement, we would see a change in the
distribution with time, the opposite of what our data show.  This
argues that the bulge growth and disk truncation should go hand in
hand.  If merging is the dominate physical process, this would imply some component of the merging 

A dynamical process is required to turn the average
star-forming galaxy into the typical early-type galaxy, for the
reasons we list above.  This process must generate a larger bulge
fraction and population with an axis-ratio distribution that is
markedly different from the flat population seen for star-forming
systems \citep[e.g.]{ryden2006,padilla2008}.  The most likely mechanism
is merging, as merging changes the axis-ratio of galaxies with low gas
masses.  Some combination of major merging and minor merging, with
more emphasis on the later due to its larger frequency, is the most
likely culprit for structurally transforming active star-forming galaxies into the
passively evolving galaxies we observe both today and at $z\sim1$.  

\subsection{Future Directions}

Our study uses a simple measurement (the projected axis-ratio) to come
to arrive at far-reaching conclusions about the evolution of galaxy
structure. The caveat is that we rely on the assumption that flattened
systems have significant rotational support.  It also rests on the
assumption that one number to characterize the intrinsic shape is a
sensible approximation, allowing us to bypass bulge-disk
decompositions \citep{macarthur2008,laurikainen2010,simard2011} and
spatially resolved, stellar dynamics
\citep{vandermarel2007a,vandermarel2007b,krajnovic2008,vanderwel2008b},
which are notoriously difficult at high redshift.  So far, such
studies support our conclusions, most explicitly by the observation
that the fraction of rotationally supported early-type galaxies is
similar at $z\sim 1$ and in the present-day universe
\citep{vanderwel2008b}.

An interesting question is whether the absence of a significant
population of very massive disk-like galaxies at $z\lesssim 1$ is a
fundamental feature of galaxy formation. Perhaps under circumstances
that are met at much earlier epochs than $z\sim 1$ such galaxies can
and do exist, and the observations presented in this paper merely show
that merging, either minor or major, is the only relevant mechanism to
produce very massive galaxies at relatively recent
epochs. Observations of significantly large samples of very massive
galaxies at $z\sim 2$ may provide answer and do show some hints
\citep{vanderwel2011}. The structural properties of galaxies at the
epoch during which the star formation rate was highest will tell us
whether galaxies with stellar masses $M>2\times 10^{11}\ M_{\sun}$ are
always bulge dominated.

\subsection{Conclusions}

In this paper we analyze the projected axis-ratio distributions of
early-type galaxies with stellar masses $>1.25\times10^{10}\ M_{\sun}$
at $z\sim 0.06$ and $z\sim 0.7$.  By modeling the intrinsic
distribution, we find that at least since $z\sim 1$, there is a
stellar mass ceiling for flattened early-type galaxies.  Above
$10^{11}\ M_{\sun}$ such galaxies are increasingly rare, both at the
present day and at $z\sim 0.7$ (see Figures \ref{rixtwo}, \ref{dist}
and \ref{mf}).  This suggests that at all cosmic epochs the dominant
evolutionary channel for early-type galaxies with higher masses is a
dynamical process that transforms systems with a 1:3 intrinsic
axis-ratio into a rounder, triaxial system with a roughly 2:3
axis-ratio.

Below that mass threshold, the early-type galaxy population becomes
more and more dominated by flattened or disk-like systems, with
roughly an axis-ratio of 1:3.  This is manifest in the number of round
galaxies as well as in the larger and larger number of galaxies that
appear thin in projection.  This geometric picture also fits very well
with the kinematic evidence that shows that most such early types are
'rapid rotators', at least in the present-day universe.  Once again,
the axis-ratio distribution appears to evolve little out to $z\sim1$.
The non-evolving shape of the mass function of flat versus round
galaxies we find in Figure \ref{mf} coupled with the overall growth of
the normalization implies that $M^{\star}$ early-type galaxies form in
a similar way over the last 7 Gyrs.  The growth mechanism must roughly
double to triple the number of early-type galaxies, producing a mix
bulge-to-disk ratios that varies with galaxy mass, but the process
must not vary with time in the mass range we study.  The leading
puzzle for early-type formation is a unifying model for how to explain
this growth in mass density with so little change in the shapes of
galaxies over the same look-back time.
\\
\\
The authors would like to thank Erik Bell, Dan McIntosh, Greg Rudnick,
Sandra Faber and David Koo for useful discussions, comments and
feedback.  BPH would also like to thank the scientists and staff of
the Max Planck Institute for Astronomy in Heidelberg for hosting him
while working on this project.

{\it Facilities:} \facility{HST (ACS)} \facility{Sloan}

\end{document}